\documentclass[twocolumn,PRA,preprintnumbers,superscriptaddress,amsmath]{revtex4}
\usepackage[latin9]{inputenc}
\usepackage{amsmath}
\usepackage{amssymb}
\usepackage{graphicx}
\usepackage{esint}

\makeatletter
\@ifundefined{textcolor}{}
{%
 \definecolor{BLACK}{gray}{0}
 \definecolor{WHITE}{gray}{1}
 \definecolor{RED}{rgb}{1,0,0}
 \definecolor{GREEN}{rgb}{0,1,0}
 \definecolor{BLUE}{rgb}{0,0,1}
 \definecolor{CYAN}{cmyk}{1,0,0,0}
 \definecolor{MAGENTA}{cmyk}{0,1,0,0}
 \definecolor{YELLOW}{cmyk}{0,0,1,0}
}

\makeatother

\begin{document}

\title{Transient Microcavity Sensor}
\author{Fang-Jie Shu}
\address{Key Laboratory of Quantum Information, University of Science and
Technology of China, CAS, Hefei, Anhui 230026, China}
\address{School of Physics and Electrical Information, Shangqiu Normal University, Shangqiu, Henan 476000, China}
\author{Chang-Ling Zou}
\address{Key Laboratory of Quantum Information, University of Science and
Technology of China, CAS, Hefei, Anhui 230026, China}
\address{Synergetic Innovation Center of Quantum Information \& Quantum Physics, University of Science and
Technology of China, Hefei, Anhui 230026, China}
\email{clzou321@ustc.edu.cn}
\author{\c{S}ahin Kaya \"{O}zdemir}
\address{Electrical and Systems Engineering Department, Washington University, St. Louis. MO 63130, USA}
\author{Lan Yang}
\address{Electrical and Systems Engineering Department, Washington University, St. Louis. MO 63130, USA}
\author{Guang-Can Guo}
\address{Key Laboratory of Quantum Information, University of Science and
Technology of China, CAS, Hefei, Anhui 230026, China}
\address{Synergetic Innovation Center of Quantum Information \& Quantum Physics, University of Science and
Technology of China, Hefei, Anhui 230026, China}

\begin{abstract}
A transient and high sensitivity sensor based on high-Q microcavity
is proposed and studied theoretically. There are two ways to
realize the transient sensor: monitor the spectrum by fast scanning
of probe laser frequency or monitor the transmitted light with fixed
laser frequency. For both methods, the non-equilibrium response not
only tells the ultrafast environment variance, but also enable higher
sensitivity. As examples of application, the transient sensor for
nanoparticles adhering and passing by the microcavity is studied.
It's demonstrated that the transient sensor can sense coupling region,
external linear variation together with the speed and the size of
a nanoparticle. We believe that our researches will open a door to the
fast dynamic sensing by microcavity.
\end{abstract}
\maketitle

\section{Introduction}

Recently, whispering gallery microcavities have stimulated a variety
of applications, including sensor \cite{Vollmer2012,Carrier2014,14Yang,14OZ,14bb,14Sfs,14EPJST,15rv,Su2015Label,grimaldi2015flow},
microlaser \cite{92APL,14Sfl}, frequency comb \cite{07N,Kippenberg2011}, optomechanics
\cite{08S,12SD,15yy}, and cavity quantum electrodynamics \cite{Aoki2006,Shomroni2014}. Because of the very
high Q-factor to mode volume ratio, light in the microcavity would
generate very strong local electromagnetic field and enhance the light-matter
interaction. Therefore, a perturbation of environment, such as refractive
index change \cite{05APL,14EPJSTCh}, tiny nanoparticle binding/debinding \cite{08PNAS,11NN,11PNAS},
and variation of temperature \cite{11OE,14OLPackaged} or pressure \cite{07JOSA,12OLS},
will changes the properties of the optical resonance. For low input-power
where the nonlinear effects can be neglected, the responses of cavity
to external perturbation are frequency shift and linewidth change.
To sense the perturbation, probe light are coupled to the cavity through
free-space or near field coupler, and the Lorentz lineshapes in transmittance
are monitored when sweeping the laser frequency through the resonance.
From the change of central frequency and linewidth, and given the
properties of the material of the microcavity, we can tell how the
environment changes with high sensitivity.

Currently, most microcavity sensors are based on probing the steady
state transmission or reflection spectra by sweeping the frequency
of probe laser slowly. Because the scanning speed (in unit of $\mathrm{MHz}/\mu\mathrm{s}$)
usually below the character speed ($4\kappa^{2}$, where $\kappa$
is the decay rate) of a microcavity, the former steady state assumption
works well. Nevertheless, the sensor depended on steady state is limited
by its low temporal resolution \cite{12OL}, so it cannot detect high
speed dynamic procedure directly. For instance, thermal dissipation
in a microcavity is realized as a phenomenon happening in $0.1\sim1$
ms which breaks the steady state assumption \cite{04OE,14APL}. Moreover,
a ringing transmission spectrum caused by a high scanning speed laser
can be used in experiment to distinguish over coupling case from the
under coupling one \cite{09COL,14SR}. Furthermore,
binding/unbinding events included in sensing of nanoparticle are transient processes themselves which arouses ringing
transmission
spectrum too \cite{11PNAS}. To the best of our knowledge, there is
a lack of a general research focusing on the transient response of
a microcavity in sensing.

In this paper, we theoretically study the transient response of high-Q microcavity.
Meanwhile, two transient sensing schemes relying on the temporal
response of microcavity is proposed, their properties and limitations are also studied.
As an examples of application, the nanoparticle motion sensors based on microtoroid
cavity is studied, with two different scenarios, i.e. particle adhering
to the cavity wall and particle passing by it. The transient response
studied here is not restricted to the WGM. The principle can also
be applied to other types of microcavities, such as photonic crystal
cavity \cite{Wang2012} and Fabry-Perot cavity \cite{Pevec2011}.

\section{Model of Cavity Spectrum}

\begin{figure}
\centering
\includegraphics[width=6cm]{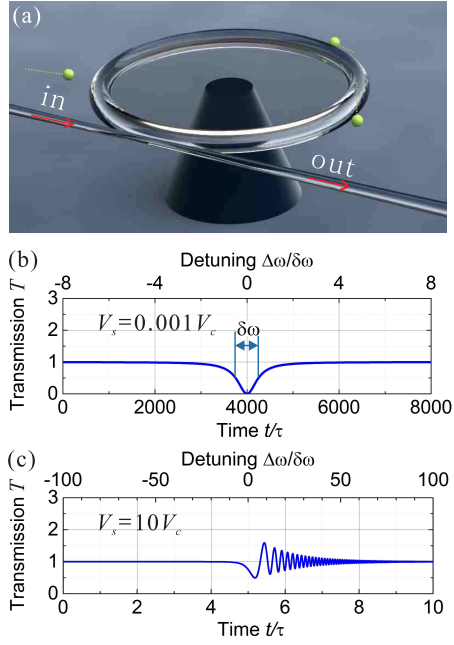}
\caption{\label{fig:A-microcavity-sensor.} (Color online) (a) Schematic illustration
of microcavity sensor for nanoparticle detection. A silica microtoroid
is coupling with a silica fiber taper, with laser being loaded from
one port and transmitted light being collected from the other. (b)
Typical quasi-static transmission spectrum of whispering gallery mode,
with the laser frequency sweeping speed $V_{s}=0.001V_{c}$ where
the character sweeping speed is $V_{c}=4\kappa^{2}$.
(c) A transient transmission spectrum for fast scanning laser, with
$V_{s}=10V_{c}$. Time normalized with $\tau=\frac{1}{\kappa}$.}
\end{figure}

Figure \ref{fig:A-microcavity-sensor.}(a) shows a typical microcavity
sensor composed of a toroid microcavity and a tapered fiber \cite{09NP,11PNAS}. A probe light is guided in the single
mode tapered
fiber, and coupled with the high-Q whispering gallery modes (WGMs)
in the microcavity through evanescent field.
When there is variation of the environment, such as the nanoparticle
approaching or leaving the microcavity, the intrinsic loss and resonant frequency
of the WGMs will be disturbed. Then, the phase and amplitude of the
cavity field will change accordingly.

The electric field component of input light generally reads
$s_{in}(t)=S_{in}e^{-j\phi(t)}$, where $j$ is the square root of
$-1$. The phase $\phi(t)$ is the integral of instant angular frequency
$\omega(t)$ about time $t$, i.e., $\phi(t)=\int\omega(t)dt$ . The
amplitude $a(t)$ of WGM field can be solved by the coupled mode theory
\cite{book,08JOSAB,99JOSAB}
\begin{equation}
\frac{da(t)}{dt}=-j\omega_{i}a(t)-(\kappa_{i}+\kappa_{e})a(t)+\sqrt{2\kappa_{e}}s_{in}(t).\label{eq:coupling}
\end{equation}
Here, $\omega_{i}$ and $\kappa_{i}$ are the angular frequency and
intrinsic loss rate of photons, which are determined by the material
and the geometry of the microcavity. The $\kappa_{e}$ is external
loss of photons cause by the fiber taper, which can be controlled
by adjusting the relative location of fiber taper in respect to the
microcavity. Denote the total decay rate of the field amplitude as
$\kappa=\kappa_{i}+\kappa_{e}$, and hence the total quality factor
$Q=\omega_{i}/2\kappa$. Moreover, the total amplitude lifetime of
photons in the mode is $\tau=1/\kappa$.

In experiments, the laser amplitude is fixed to reduce the noises
induced by amplitude fluctuation in sensor application, thus $S_{in}$
is a constant. For monochrome light with fixed laser frequency, $\phi(t)$
increases with time $t$ linearly, i.e. $\phi(t)=\omega_{in}t$ ,
we obtain the steady state solution as
\begin{equation}
a(t)=\frac{\sqrt{2\kappa_{e}}S_{in}e^{-j\omega_{in}t}}{j(\omega_{in}-\omega_{i})+(\kappa_{i}+\kappa_{e})}.\label{eq:Tss}
\end{equation}
The cavity field approaches the steady state with the build-up time
in the order of $\tau$. For the WGM, there is another time scale
$\tau_{c}$, which is the round trip time corresponding to the period
that a pulse travels around the perimeter of a microcavity. In high
$Q$ microcavity, the $\tau\gg\tau_{c}$ is satisfied, which implies
that photons travel thousands of times in the microcavity before they
lost. The high sensitivity of microcavity sensor benefits from this
repeated travels. Furthermore, we assume the time scales of all interactions
is much longer than the $\tau_{c}$, so the Eq.$\,$(\ref{eq:coupling})
is always true.

In practice, the transmission spectrum are obtained by sweeping the
frequency of the probe laser. For example, a linear frequency scanning
$\omega(t)=\omega_{in}+V_{s}t$ (in the following, the scanning speeds
is defined in terms of the angular frequency) is usually employed,
then
\begin{equation}
\phi(t)=\int_{0}^{t}\omega(t')dt'=\omega_{in}t+\frac{V_{s}}{2}t^{2}.\label{eq:Linear}
\end{equation}
A general solution of
Eq.$\,$(\ref{eq:coupling}) reads
\begin{align}
a(t)= & \sqrt{2\kappa_{e}}S_{in}e^{j\omega_{i}t-\kappa
t}\left[\frac{\tau}{1+j(\omega_{in}-\omega_{i})\tau}\right.\nonumber \\
 & \left.+\int_{0}^{t}e^{j\phi(t')-j\omega_{i}t'+\kappa t'}dt'\right].\label{eq:solution}
\end{align}
So long as we know the function $\phi(t)$, the cavity field $a(t)$
can be solved accurately by Eq.$\,$(\ref{eq:solution}). For the
linear scanning case {[}Eq.$\,$(\ref{eq:Linear}){]}, the equation
can be solved analytically \cite{14SR}. For more general cases with
nonlinear scanning, $a(t)$ should be solved numerically.

The transmission signal
\begin{equation}
s_{out}(t)=s_{in}(t)-\sqrt{2\kappa_{e}}a(t)\label{eq:input-output}
\end{equation}
corresponding to the interference of the directly-transmitted input
laser and cavity output in the tapered fiber. The intensity of transmitted
signal ($|s_{out}(t)|^{2}$) is collected and monitored by a high-speed photon detector in real-time.
For convenience, people measure the intensity spectrum of output signal
normalized by input signal
\begin{equation}
T(t)=|s_{out}(t)/s_{in}(t)|^{2}\label{eq:T(t)}
\end{equation}
to gather information from spectrum of intensity, but the phase information cannot be obtained without interferometer
which requires complex stabilization equipments.

\begin{figure*}
\centering
\includegraphics[width=16cm]{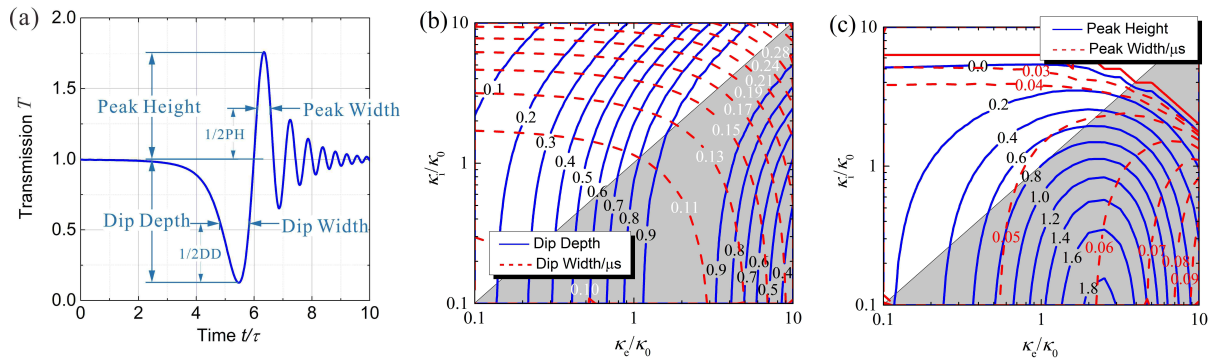}
\caption{\label{fig: fig2} (Color online) (a) Typical transmission against
time of transient sensor at critical coupling, for $\kappa_{i}=\kappa_{e}=\kappa_{0}$
with $\kappa_{0}=3.0\:\mathrm{MHz}$. (b) and (c) are diagrams for
checking coupling parameters by the values of dip width and depth
or the peak width and height. The vertical coordinate is $\kappa_{i}/\kappa_{0}$,
and the abscissa is $\kappa_{e}/\kappa_{0}$ in logarithm scale with
$\kappa_{0}$. The diagonal dash line denotes critical coupling condition
for steady state. Areas under (shade triangle) and upper the diagonal
line correspond to over- and under- coupling regimes, respectively.}
\end{figure*}

Here, we introduce the character speed $V_{c}=4/\tau^{2}=4(\kappa_{i}+\kappa_{e})^{2}$,
which corresponds to the speed that the laser sweep through the cavity
resonance (frequency range $2/\tau$) within the lifetime of the mode
$(2\tau)$. When the scanning speed $V_{s}=0.001V_{c}\ll V_{c}$,
the transmission shows the typical symmetry Lorentz spectrum, as shown
in Fig.$\,$\ref{fig:A-microcavity-sensor.}(b) for critical coupling
where $\kappa_{i}=\kappa_{e}$. In this case, the transient response
of the WGM can be neglected and the field amplitude can be approximated
by the steady state solution (Eq.$\,$(\ref{eq:Tss})). However,
in Eq.$\,$(\ref{eq:T(t)}) the information about the phase of the
transmitted light is lost, and we cannot distinguish between the under-
and over-coupling since the roles of $\kappa_{i}$ and $\kappa_{e}$ in Eq.$\,$(\ref{eq:Tss}) are identical.
Then, we cannot distinguish the
change of the intrinsic cavity loss due to the sensing event from
the change of the external cavity loss due to vibration of tapered
fiber.

For fast sweeping speed that $V_{s}$ is comparable with or even larger
than $V_{c}$, the transient response of the cavity field cannot be
omitted. For instance, the spectrum of $V_{s}=10V_{c}$ is shown in
Fig.$\,$\ref{fig:A-microcavity-sensor.}(c), where other parameters
are the same with Fig.$\,$\ref{fig:A-microcavity-sensor.}(b). As
firstly pointed out in Ref.$\,$\cite{08JOSABYDumeige}, the information
of $\kappa_{i}$ and $\kappa_{e}$ can be extracted from the transient
response deterministically. It can be understood intuitively that
the coupled microcavity and fiber taper system becomes a heterodyne
interferometer for transient sensor. Due to the high-Q WGM, there
is a hysteresis between the input light and cavity output light, or we can
say the light emitted from the cavity is actually the light input from the fiber
a while ago. When $V_{s}$ is large, the detected light is the interference
between lights of different frequencies, thus it contains both the
amplitude and phase information of the cavity field.

\section{Transient Response of Microcavity}

In general, there are two ways to monitor the variation of cavity
resonance. One is sweeping the laser frequency quickly and repeatedly.
The frequency $\omega(t)$ is a linear function with time, then the
phase $\phi(t)$ is a quadratic function of time. Since each frequency
sweeping takes a repetition time $P$, we can refresh the information
of environment at digital time $nP$ ($n$ is an integer). The other
approach is locking the laser frequency precisely, and then monitoring
the change of transmitted signal due to the change of resonant frequency.
In the later case, the relevant $\phi(t)=\int\omega(t)dt$ is very complicated, which is determined by the history of the
variation of environment. For
both cases, the principle
and applied theory are the same with those described in previous section.

\subsection{Sweep Laser Frequency}

The resonant frequency of a cavity can be characterized by monitoring
the position of dip in transmission spectrum of steady state in experience.
In addition, relying on the transient response analysis, people has
figured out how to get $\kappa_{e}$ and $\kappa_{i}$ from the parameter
fitting method \cite{08JOSABYDumeige,09COL}. Here, we provide a direct
way to determine the coupling rates by the features of transmission
curves $T(t)$ in transient states. To show the results quantitatively,
we choose the following parameters which are reasonable in experiments:
resonant wavelength $\lambda_{i}=1.55\ \mu\mathrm{m}$ and $Q_{i}=2\times10^{8}$.
The critical sweep speed for such high-Q cavity with critical coupling
is $V_{c}=4(\frac{\omega_{i}}{Q_{i}})^{2}\approx148\ \mathrm{MHz}/\mu\mathrm{s}$.
When the frequency of the input signal scans fast (e.g. $V_{s}=100\mathrm{MHz}/\mu\mathrm{s}\approx0.67V_{c}$)
through a resonant frequency of the microcavity, an asymmetry dip
followed with a ringing tail is the typical shape of $T(t)$ (Fig.$\,$\ref{fig: fig2}(a)),
in contrast to a simple symmetry Lorenz lineshape dip of $T(t)$ in
steady state assumption. Because of the transient process that light builds
up inside the cavity, the dip bottom in on-resonance moment does not
reach $0$. The ringing tail is the result of interference between back coupled WGM light and the transmitted signal light. Besides,
for a system being free from nonlinear effect \cite{09COL}, if the
input signal scans downward ($-V_{s}$) through the resonant frequency,
the transmission curve is the same in time sequence as it scans upward.

The analytical solution of the integral in Eq.$\,$(\ref{eq:solution})
can be written as \cite{14SR}
\begin{equation}
a(t)=\sqrt{2\kappa_{e}}S_{in}e^{j\omega_{i}t-\kappa t}\left[f(t)-f(0)+\frac{1}{\kappa+j(\omega_{in}-\omega_{i})}\right],
\end{equation}
where
\begin{equation}
f(t)=-\sqrt{\frac{j\pi}{2V_{s}}}e^{-\frac{[j(\omega_{in}-\omega_{i})+\kappa]^{2}}{2V_{s}}}\mathrm{erf}(\frac{j\kappa+\omega_{i}-\omega_{in}-V_{s}t}{\sqrt{2jV_{s}}})
\end{equation}
and $\mathrm{erf}(z)$ is the complex error function.

The transmission curve can be characterized by four parameters: peak
height, peak width, dip depth and dip width {[}Fig.$\,$\ref{fig: fig2}(a){]},
where peak width is the full width of half maximum of the first peak,
dip width is the full width of half minimum of the first dip. Varying
$\kappa_{i}$ and $\kappa_{e}$ and fixing $V_{s}=2V_{0}\approx296\ \mathrm{MHz}/\mu\mathrm{s}$,
these parameters are extracted, and are plotted in Figs.$\,$\ref{fig: fig2}(b)
and (c). Noting that the absolute position of the dip is not shown here, which corresponding to the shift of cavity
resonance, because it can be determined in experiment precisely with the help of another
reference cavity.

If both the $\kappa_{i}$ and $\kappa_{e}$ are large enough to make
$V_{s}\ll V_{c}=4(\kappa_{i}+\kappa_{e})^{2}$ (top right of Fig.$\,$\ref{fig: fig2}(b)),
it is the quasi-steady state where the dip depth depends on the ratio
$\kappa_{i}/\kappa_{e}$. When $\kappa_{i}\approx\kappa_{e}$,
the dip depth is around $1$ corresponding to the critical coupling.
However, when decreasing the $\kappa_{i}$ and the $\kappa_{e}$,
the critical coupling deviates from the line $\kappa_{i}=\kappa_{e}$.
When $\kappa_{i}$ is small (lower half of Fig.$\,$\ref{fig: fig2}(b))
the dip depth mainly depends on $\kappa_{e}$ , and
the extreme value of dip depth appears near the $\kappa_{e}\approx2\kappa_{0}$.
It is because that when $V_{s}>V_{c}$, the critical coupling for
transient response requires $\kappa_{e}>\kappa_{i}$ to make energy enter
the cavity more efficiently and afterwards destructively interference with the input light, leads to a cancelation of transmission.
However, if the $\kappa_{e}$ is too large, the energy of cavity couples back to tapered fiber
faster, thus break the balance and give non-zero transmission. So, as analogue to the steady state critical coupling, the
certain $\kappa_{e}$ give rise to temporal critical coupling that dip depth be unitary.

In both Figs.$\,$\ref{fig: fig2}(b) and (c), in the transient region
that $V_{s}\geq V_{c}$, the contours of the width and height of peak
and the width and depth of dip cross, respectively. Therefore, the
transient transmission provides extra information from which we can
determine the value of $\kappa_{i}$ and $\kappa_{e}$. For example,
if a coupling system has a transmission curve with a dip depth of
$0.8$ and dip width of $0.13$, it may in over or under coupling
region (see Fig. \ref{fig: fig2}(b)). Then check the peak width, assume $0.07\ \mu\mathrm{s}$
to obtain the value of $\kappa_{i}\approx3.2\kappa_{0}$ and $\kappa_{e}\approx4.4\kappa_{0}$.
There even exists other redundant feature (peak height) to double
check the result. Therefore, combining with the measured location
of the resonance, we can estimate the instant frequency and decay
rate of the WGM for sensing purpose.

More generally, for any given $V_{s}$, we can normalize the transmission
$T(t)$ (Eqs.$\,$(7) and (8)) by replacing $\kappa/\sqrt{V_{s}}$
with $\kappa'$ and $t\sqrt{V_{s}}$ with $t'$, respectively. Then
the $\kappa_{i}$, $\kappa_{e}$, and $\omega_{i(in)}$ are rescaled
with a factor of $1/\sqrt{V_{s}}$, and replaced with corresponding
prime variables too. After the transformation, $T$ is independent
of $V_{s}$, so the charts of Figs.$\,$2 (b) and (c) are sufficient
to estimate the $\kappa_{i}$ and $\kappa_{e}$ at any really applied
sweep speed $V_{s}$. For example, if the sweep speed is $V_{s}=8V_{0}$,
the measured peak/dip width should be rescaled with a factor of $2$.
Using the rescaled widths to find out the apparent $\kappa_{i(e)}$
from the Figs.$\,$\ref{fig: fig2}(b) and (c), and rescale back the
result $\kappa_{i(e)}$ with a factor of $2$ to get the real $\kappa_{i(e)}$
at current sweep speed.

\subsection{Fixed Laser Frequency}

For the other method, a coupled microcavity and tapered fiber system
can be calibrated and stabilized with fixed coupling and loss rates,
and the frequency of the incident probe light is locked to the cavity
resonance. The environment variation will change the dielectric properties
of the cavity and cause a change of $\omega_{i}$, then the detuning
$\Delta\omega$ between probe light and cavity mode $\omega-\omega_{i}$
occurs, resulting in the change of transmittance. It has been proved
in Fabry-Perot cavity that the linear increase of $\omega_{i}$ is
equivalent to the decrease of $\omega$ with similar scanning speed
\cite{02AO}. Therefore, a persistent changing of $\omega_{i}$ at
relatively high speed can generate a transient ringing transmission
line too.

\begin{figure}
\centering
\includegraphics[width=5.2cm]{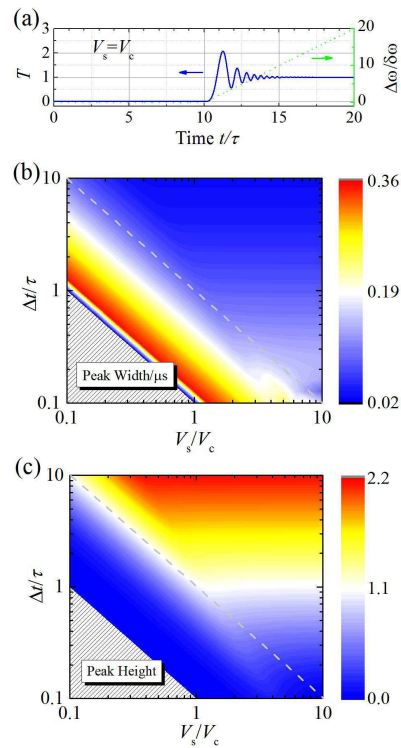}
\caption{\label{fig3} (a) The transmission response based on linear variation
of resonance frequency beginning with critical coupling. The peak
width (b) and height (c) varied with resonance sweeping speed and
elapse time.}
\end{figure}

Taking a linear change of $\omega_{i}$ \cite{11APL} and initial
critical coupling as an example, we investigate the feature of the
transient transmission. For a general consideration, we study the spectrum for various cavity frequency change
frequency and dynamic range, thus solve the temporal transmission numerically.
A typical transmission curve is shown in Fig.$\,$\ref{fig3}(a).
In this example, the change of $\omega_{i}$ begin at time $10\tau$.
The transmittance is zero when $t<10\tau$ because the system state
is kept in critical coupling and the probe light is absorbed by the
cavity completely. When $\omega_{i}$ start to change due to environment
perturbation, the balance between the input laser and cavity leakage
is broken suddenly, then a ringing line shape is formed. This scenario
is similar as the case studied in previous subsection, where the laser
frequency is sweeping while the cavity frequency is fixed. But,
when the sweep speeds are the same the peak height is higher now than
that in the sweep laser case.

In practice, $\omega_{i}$ may drifts in a period of time $\Delta t$
at sweep speed $V_{s}$, then reaches the equilibrium position with the final detuning $\Delta\omega=V_{s}\Delta t$.
Then, the ringing tail will end to the steady state transmission
level $T_{s}=(\triangle\omega)^{2}/(1/\tau^{2}+(\triangle\omega)^{2})$,
which may less than unit evidently. Correspondingly the definitions
of peak height and peak width of $T(t)$ are modified according to
the horizon line of $T=T_{s}$, as compare to Fig.$\,$\ref{fig: fig2}(a).
Peak width and peak height vary with $\Delta t$ and $V_{s}$ (Figs.$\,$\ref{fig3}
(b) and (c)). In those figures of double logarithmic coordinates,
$\Delta\omega$ on each of lines paralleling the diagonal dash line,
on which $\Delta\omega$ equals two times of line-width, are constant.
In the lower left of the figures, shaded triangle infers there is
no peak higher than $T_{s}$. However, in this regime, the total frequency change is still within the cavity resonance and the change of cavity resonance is slow, the environment change can be estimated through quasi-static cavity response.
 At the right vertex of the triangle,
though the sweep speed is as fast as critical speed, the short elapse
time $0.1\tau$ makes a narrow detuning $\Delta\omega\approx0.2\times(2/\tau)$.
It means the $T_{s}\approx0.14$, and in the time of sweeps, transmission
is kept at the relative flat bottom of Lorentz line-shape. The relative
gentle variation of the transmission restrains the ringing phenomena.
When the $\Delta\omega$ gets larger, first a broaden low peak arises,
then it becomes sharper and higher.

Furthermore, if the $\Delta\omega$ is larger than two times of line-width
(right upper region of the figures), the peaks are no longer varied
with $V_{s}$, especially in high speed side. This saturation about
speed can be understood as the response time of the microcavity is
limited to life time $2\tau$. Though the $\omega_{i}$ drifts beyond
the resonant frequency in a short time far less that $2\tau$ at high
sweep speed, the energy in the microcavity leaks out the cavity with
its own pace. In addition, comparing the saturated peak height on
the most right vertical line with the color bar, we can find peak
height is proportional to $\log\Delta t$.

\subsection{Comparison}

For the two different approaches, we can find there are many differences.
In the sweeping laser case, the frequency range of the sweeping is
usually in the order of $2\pi\times10\sim100$ GHz, which provide
a wide dynamic range of the sensor. However, for a fixed sweeping
speed $V_{s}$, the larger dynamic range also means longer repetition
time $P$ for the laser sweeping through the wide frequency range.
For example, the frequency of WGM in the silica microcavity may change
by $2\pi\times1.7$ GHz when the environment temperature changes by
$1\ \mathrm{K}$ \cite{11OE}. If the sweeping speed $V_{s}=2\pi\times200\ \mathrm{MHz}/\mu\mathrm{s}$
and frequency range is $2\pi\times200$ GHz, the dynamic range of
the temperature sensor is $117$ K while $P=1\ \mathrm{ms}$. If the
frequency range reduced by $100$ times, the dynamic range will only
be $1.17$ K while $P=10\ \mu\mathrm{s}$. Therefore, there is a
trade-off between the temporal resolution and dynamic range. Note
that the sweeping laser transient sensor pre-assumed that the environment
induced change speed of frequency or decay rate is much slower than $V_{s}$.
For a given $V_{s}$, higher Q gives higher sensitivity.

In contrast, the fixed frequency case show a much smaller dynamic
range, which can only sense the event with frequency shift smaller
than the linewidth, and is usually several orders of magnitude smaller than that in the sweeping
laser case. The benefit is that this type of transient sensors can
response to the external perturbation quickly within the time-window of $1\sim100\,\mathrm{ns}$,
which is $4-6$ orders smaller than $P=1\ \mathrm{ms}$ in the sweeping
laser case. For example, a nanoparticle pass through the cavity may
induce a small frequency change within $1\ \mu\mathrm{s}$. This
event may be missed for the sweeping laser case, but can be monitored
by the fixed laser case. There is a trade-off between the dynamic
range (linewidth of resonance) and response time (lifetime of resonance),
and the sensitivity is also related to the linewidth.

Therefore, the two different transient sensors are appropriate for
different applications. When the perturbation induced frequency shift
is very large but relatively slow, we can use the sweeping laser.
When the perturbation is small but very fast, we can use the fixed
laser. In both cases, an external reference cavity is required, which
is used as frequency reference in the laser sweeping case or to lock
the laser frequency for the fixed laser frequency case.

It's also worth to note that the sensitivity of the transient sensor
can also be improved comparing with that in the slowly sweeping case.
In the transient response, the ringing transmission curve is sharper
due to increased slope of $dT/d\omega$ (the time is usually converted
to frequency by $V_{s}$ in experiment). For example, the maximum
slope (the first ring up) for $V_{s}=2V_{c}$ is about 4 times larger
than the maximum slope for $V_{s}\ll V_{c}$. In addition, the refresh
rate is also increased by increasing $V_{s}$. One may argue that
the shot noise of the laser also increase for faster sweeping laser,
since the duration of time to sweep through the cavity resonance is
proportional to $1/V_{s}$. This can be improved by increasing the
laser power for sweeping, the only limitation is that the power absorbed
by the cavity should be smaller to avoid thermal or Kerr effect. Similarly,
the slope will be larger for faster perturbation in the fixed laser
frequency case, where the sensitivity depends on sensing event but
not the equipments in experiments.

\section{Example: Transient Nanoparticle Sensor}

In this section, we will analyze the practical application of the
microcavity sensor for transient nanoparticle motion. The nanoparticle
sensor can be used for study the kinetic properties of the nanosized
particles, with micrometer-scale spatial resolution, and microsecond-scale
temporal resolution. Use the setup demonstrated recently \cite{11NN},
we can explore the mechanics, fluid dynamics and stochastic Brownian
motion with unprecedented precision. When a nanoparticle approaches
the microcavity, it shifts the resonance ($\omega_{i}$), and also
increases the loss of the microcavity by scattering. But in high $Q$
microcavity, the internal lifetime $\tau_{i}$ mainly depends on the
absorption loss of the cavity material. Scattering loss like binding
a nanoparticle or carving a small notch does not affect the lifetime
much \cite{10PNAS}. So we can safely ignore the change of the $\tau_{i}$,
and fix the system in critical coupling region. In the following,
we study the transient sensor initialized in critical coupling region,
when the particles are far away.

\subsection{Binding }
\begin{figure*}
\centering
\includegraphics[width=16cm]{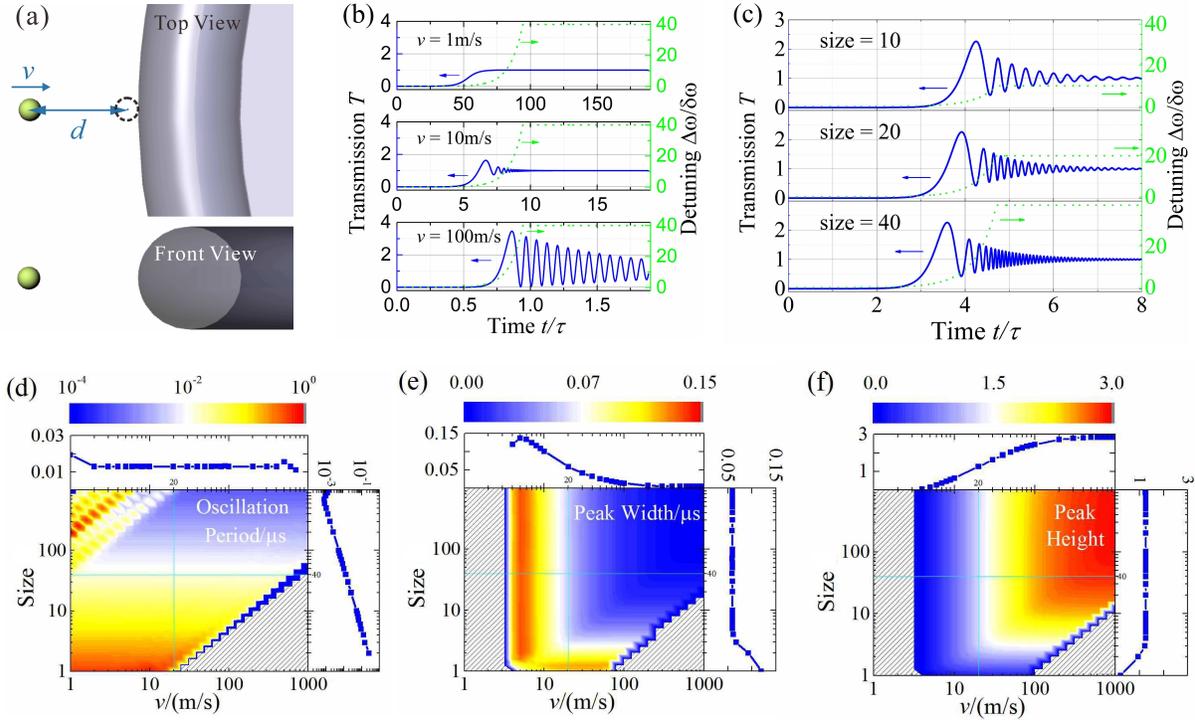}
\caption{(a) Schematic illustration of the binding nanoparticle sensing, where a particle approaches
to the side of cavity and then adhere on it; (b) and (c) the transmissions $T(t)$ for fixed probe laser frequency for different particle traveling speed ($v$) and sizes ($\epsilon$). (d)-(f) The features of the transmission versus $v$ and $\epsilon$.}
\label{fig4:binding}
\end{figure*}
As illustrated in Fig.$\,$\ref{fig4:binding}(a), a nanoparticle
approaches the microcavity along a trace perpendicular to the outer
rim, then is bound \cite{Su2015Label} on the surface . The nanoparticle shifts the frequency
of a mode by moving into the evanescent field of the mode. The degree
of the shift is proportional to local field intensity \cite{03OL}.
Besides, the evanescent field of a mode decays (decay factor $\alpha$)
exponentially along with the distance $d$. Therefore, a nanoparticle
approaching the microcavity with constant velocity $v$ will induce a mode frequency shift which is exponentially increase with time (Figs.$\,$\ref{fig4:binding}(b) and (c)
the first half of dotted green line)
\begin{equation}
\omega(t)=\omega_{i}(1-\epsilon e^{-\alpha(d_{0}-vt)}).\label{eq:exp}
\end{equation}
Here for the convenience of symbol system, the variation of $\omega_{i}$
has been equivalently transferred to $\omega(t)$. At time $t=0$,
the nanoparticle is in distant $d_{0}$ ($\alpha d_{0}\gg1$), and
$\omega(0)\approx\omega_{i}$. When the nanoparticle touches the microcavity
and be adhered to it, the $\omega(t)=\omega(d_{0}/v)$ for $t>d_{0}/v$
is a constant (the last half  of
dotted green line in Figs.$\,$\ref{fig4:binding}(b) and (c)). The $\epsilon$, depending on inherent electromagnetic
attributes of the particle, represents the degree of frequency shifting
when the particle is bound on the surface. Replace the $\omega(t)$
of Eq.$\,$(\ref{eq:solution}) with the Eq.$\,$(\ref{eq:exp}),
we can get the transmission $T(t)$ numerically.

Set $\epsilon=size/Q=40/Q$, which means the maximum shift $\epsilon\omega_{i}$ is $40$
times of linewidth. Other parameters are the same with
that in Section III. Firstly, the transmission and its features varying
with speed of the particles are investigated. The transmissions $T(t)$
of various speeds are shown in Fig.$\,$\ref{fig4:binding}(b). Similar
to the previous analysis, the transient frequency shift induces ringing
phenomena, the peak value increases with the increasing of speeds.
In addition, the transmission for different particle sizes/volumes are
studied, as shown in Fig.$\,$\ref{fig4:binding}(c). Since the total
resonant frequency shift $\epsilon$ is proportional to size/volume
of the particle, the effective frequency shifting speed increases
with the size. According to equations (6) in Ref. \cite{88OL}, the
oscillation period of ringtail is inversely correlated to $\epsilon$ and thus decreases with
increasing of size.

The basic features, including oscillation period, peak width and peak
height, varying with the particle size and speed are plotted in Figs.
4(d)-(f). Using the features measured in experiments, we can identify
the particle size and speed, since they correlate with these parameters
in quite different ways. For example, from the curves extracted from
the 2D maps, when size=40, the oscillation period is almost constant
(Fig. \ref{fig4:binding}(d) up panel), while peak width (Fig. \ref{fig4:binding}(e)
up panel) and height (Fig. \ref{fig4:binding}(f) up panel) vary;
when speed is fixed by 20 m/s, the period almost inversely proportional
to size (Fig. \ref{fig4:binding}(d) right panel), while the other
two features remains the same (Figs. \ref{fig4:binding}(e) and (f)
right panels). Although the regime for small $v$ does not show ringing phenomena, but it's quasi-static process, we can determine the size and velocity from the slowly varing transmission line. For the down-right coner in Figs. 4(e) and (f), there is also no ringing because the transient sensor is ultimately limited by the cavity lifetime.

\subsection{Passing-by}

\begin{figure*}
\centering
\includegraphics[width=16cm]{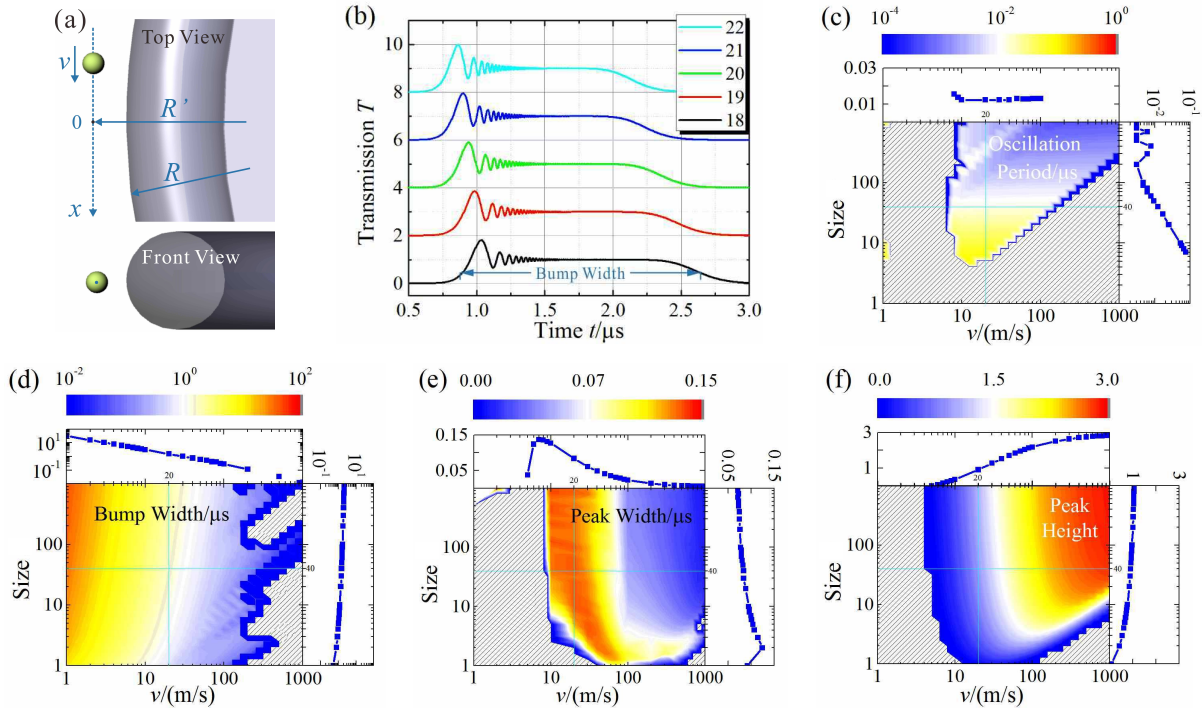}
\caption{(a)  Schematic illustration of the passing-by nanoparticle sensing . The radius of microcavity is $R$, the aim distance is $R'$. (b) The transmission $T(t)$ varies with speed. Curves are shifted up for clear show. (c)-(f) are the features: oscillation period, bump
width, peak width, and peak height versus nanoparticle size and speed.}
\label{fig5:gauss}
\end{figure*}

In another practical sensor scenario, nanoparticle passes by the microcavity
in a tangential way (Fig.$\,$\ref{fig5:gauss}(a)). Set $x$ axis
along the path. Its positive direction is the direction of velocity
of the nanoparticle. Original point $x=0$ is the point which is nearest
to the microcavity on the path. Then, the frequency shift reads
\begin{equation}
\omega(t)=\omega_{i}(1-\epsilon e^{-\alpha(\sqrt{R'^{2}+(x_{0}+vt)^{2}}-R)}).\label{eq:10}
\end{equation}
The $x_{0}(<0)$ is the initial position of nanoparticle when $t=0$.
The nanoparticle stays in the effective couple range in a short time
inversely proportional to speed (Fig. \ref{fig5:gauss}(b)). In the
$T(t)$ curve, we define bump width with full-width at $T=0.5$ to character it (Fig. \ref{fig5:gauss}(b)
button panel). The characters of the $T(t)$ concerning different
particle speeds and sizes are performed (Figs.$\,$\ref{fig5:gauss}(c)-(f)).
Like the binding case, the period is only related to the size. The
peak height, peak width and the bump width mainly vary with the speed.

At first glance, the contours of features of $T(t)$ in the case of
passing by are like that in the case of binding, except for a new
feature bump width. But in detail investigation, the variations of
features depicted in Fig. \ref{fig5:gauss} are complicated than their
counterpart in Fig. \ref{fig4:binding}. Because in content of frequency
shift, Eq. (\ref{eq:10}) is more complicated than the simple exponential
relationship depicted by Eq. (\ref{eq:exp}). The Eq. (\ref{eq:10})
is a exponential function, when $|x_{0}+vt|\gg R'$ (i.e., particle
is far away from the microcavity). While it is simplified to a Gaussian
function, when $|x_{0}+vt|\ll R'$. Then the actual frequency shift
is between the two extreme situations. Based on the Figs. \ref{fig4:binding} and \ref{fig5:gauss}, we find
that the oscillation period is suitable to characterize the particle
size in experiment. The peak width has the best sensitivity to detect
particle speed when particle size is not less than 5.

\section{Conclusions}

In summary, based on the transient transmittance of high-Q microcavities,
the environment variation can be sensed with both high sensitivity
and high temporal resolution. In addition to the enhanced frequency
shift sensing, we can also sense the intrinsic and external cavity
loss. Especially, we studied the motive nanoparticle sensing and demonstrated
that the speed and size of a nanoparticle can be determined by the
experiment-measurable transmission. We believe that our studies opens
a door to fast dynamic sensing by microcavity. Combining with the
package technique, we would expect to bring the fiber integrated optical microcavity
sensors \cite{11OE, Wang2012, Pevec2011} out of lab for ultrafast and ultrasensitive detections.

Note added: During the preparation of this manuscript, a related experiment work published \cite{Rosenblum2015},
where a cavity ring-up spectroscopy with submicrosecond time resolution by abruptly turn-on a far-detuned probe laser
pulse is demonstrated.

\section*{Funding Information}
``Strategic Priority Research Program(B)'' of the Chinese Academy of Sciences (XDB01030200); National Basic Research
Program
of China (2011CB921200, 2011CBA00200); National Natural Science Foundation of China (NSFC) (11204169); the Young
Core
Instructor Foundation from the Education Department of Henan Province, China (2013GGJS-163); Open Project of Key
Laboratory
of Quantum Information (KQI201502); Army Research Office (ARO) (W911NF-12-1-0026).

\section*{Acknowledgments}

We want to thanks Qijing Lu, Xiao Xiong for their initial contribution
to this work. C.-L.Z. appreciates the discussions with Chun-Hua Dong,
Hailin Wang and Liang Jiang.




\end{document}